# Clarifying Theoretical Intricacies through the Use of Conceptual Visualization: Case of Production Theory in Advanced Microeconomics

Alexandra Naumenko[1] & Seyyed Ali Zeytoon Nejad Moosavian[2]

[1] North Carolina State University, USA

[2] North Carolina State University, USA

Correspondence: Seyyed Ali Zeytoon Nejad Moosavian, North Carolina State University, USA

**Abstract**

Production theory, defined as the study of the economic process of transforming inputs into outputs, consists of two simultaneous economic forces: cost minimization and profit maximization. The cost minimization problem involves deriving conditional factor demand functions and the cost function. The profit maximization problem involves deriving the output supply function, the profit function, and unconditional factor demand functions. Nested within the process are Shephard's lemma, Hotelling's lemmas, direct and indirect mathematical relations, and other elements contributing to the dynamics of the process. The intricacies and hidden underlying influences pose difficulties in presenting the material for an instructor, and inhibit learning by students. Simply put, the primary aim of this paper is to facilitate the teaching and learning of the production theory realm of Economics through the use of a conceptual visual model. This paper proposes a pedagogical tool in the form of a detailed graphic illustrating the relationship between profit-maximization and cost-minimization under technical constraints, with an emphasis on the similarities and differences between the perfect-competition and monopoly cases. The potential that such a visual has to enhance learning when supplementing traditional context is discussed under the context of contemporary learning literature. Embedded in the discussion is an example of how we believe our model could be conceptualized and utilized in a real-world setting to evaluate an industrial project with an economic point of view.

## 1. Introduction

As many educators have thus far pointed out, the younger generation of college students differs from the older generation in many ways. Living in a world full of images, animations, media, and visual technologies has turned them into a new generation of visual learners. As Nilson (2010) mentions, "the younger generation of students is not as facile with text as it is with visuals, so a wise idea is to illustrate courses' designs to students so they can see where the course is going in terms of students' learning." Also, as Nilson (2010) explains, it is very unlikely that students build cognitive schemata of the material being covered in one or two semesters of casual study. Therefore, it is instructors' mission to provide their students with relevant visual big pictures that represent the structure of the associated discipline with valid, ready-made frameworks for fitting the content. She also mentions that without such a big picture, students face another learning hurdle in addition to their other hurdles they may have. In his paper entitled "Teaching Economics and Providing Visual Big Pictures", Moosavian (2016a) concludes that there is enormous potential with visualization to improve the quality of teaching and learning in economics, which has not yet been fully employed to resolve some of the issues with the teaching of economics. The present paper is an attempt to offer a remedy for the confusion that economics students may face when learning producer theory for the first time. Producer theory is comprised of two facets of optimization (cost-minimization and profit-maximization), guided by constraints, and is easy for the students to lose sight of the big picture and fail to understand how the process of producer theory operates in its entirety. This paper will present a graphic which could serve as a valuable learning tool for enhancing conceptualization and solidifying understanding.

As Nilson (2010) puts it, an instructor should contrast the effortless, holistic process of visual perception with the more draining task of learning from printed material. The nature of text is to organize and present information hierarchically, one piece at a time, so naturally, the mind processes it in the very same way – piece by piece. For this reason, learning solely through text may limit conceptualization of the big picture. Without a visual, the mind is inhibited from other productive aspects of learning, such as considering implications and relating it to other material.

That being said, the instructor is responsible for presenting the material in a way that would help the student navigate the knowledge in a way that is clear and digestible and eliminates room for misconceptualization. She endorses visuals as a means of facilitating accurate interpretation of the material and slowing down the mental facilities so that the mind has enough time to recognize and understand the patterns among concepts.

Our proposed visual aid presents the comprehensive wheel of relationships for producer theory. As far as we can tell, we are the first to offer such a visual. The graphic offers insight into the two ends of the spectrum of producer theory – perfect competition and monopoly. Though the paper will mention some theoretical aspects of imperfect competition too, our paper serves to discuss only the two extremes. The other market structures will be decomposed in future papers.

The present paper will illustrate the relationship between cost-minimization and profit maximization, bounded by technical limitations. How the relationships of the individual forces differ from those in consumer theory will be discussed. The paper will also evaluate the similarities and differences between perfect competition and monopoly. As far as we are aware, we have not seen the parametric solutions for the monopoly case which we offer in any textbooks. A practical real-world example of how both of the visuals can be used in practice will accompany our theoretical discussion. Embedded throughout the paper are suggestions for how these visuals can help practitioners effectively evaluate industrial or entrepreneurial projects through an economic lens.

The paper is organized as follows. The next section reviews the existing literature related to the role of visual models in teaching and learning. Next, the main discussion will reveal the two visual models we constructed to illustrate the wheel of relationships in production theory. The discussion will explain the models themselves as well as discuss the power these models have to facilitate teaching, learning, and application when used in conjunction with traditional economics textbooks. Naturally, a conclusion will follow the discussion section bringing the main points together and discussing plans for future research. Lastly, the paper will end with appendices to explain notation and provide numerical examples of how the two visuals can be used in practice.

## 2. Literature Review

In this section, the existing literature on the importance of visualization in teaching and learning is reviewed with an emphasis on visual "big pictures" and their potential roles in the teaching of economics. To do so, a select set of studies have been reviewed from the education literature, primarily concerning the importance of visualization and visual "big pictures" in the process of learning. The information and knowledge gained from this part will be used to build up a theoretical, educational framework to support the main purpose of the paper.

There are several major studies in the literature of teaching proving that visual displays of any form enhance learning (Vekiri 2002, Winn 1991). Quite simply, one of the benefits of incorporating visuals in instruction is being able to accommodate a wide range of learning styles. A visual which shows how concepts and processes interrelate through spatial relationships, images, colors, and codes allows the instructor to reach his or her visual, global, and concrete learners, who prefer to think in pictorial, spatial, and sensate terms (Clark and Paivio 1991, Fleming and Mills 1992, Svinicki 2004, Theall 1997). Thus, the visual serves as a means for deeper understanding by the student, enhancing their comprehension.

Visuals also have the added benefit of "perceptual enhancement." The spatial organization of models allows the learner to process the components as a whole all at once (Larkin and Simon 1987). This feature of visual aids illuminates the nature of the material, and thus, results in a deeper and more satisfying grasp of the concept or discipline. Visuals enhance learning by transparently displaying the spatial arrangements, directions of arrows, and other graphic features which allow the learner to distinguish causal links and directions, strength of relationships, and the general nature of the subject (Nilson 2009).

In her prominent book "Teaching at its Best," Nilson (2010) considers the relationship between reading and writing text and conceptualizing the information. She suggests that failure of the instructor to provide a visual aid can result in inaccurate retrieval of concepts. Nilson argues that students should be considered disciplinary novices. Since they are inexperienced, incomplete information could result in misconceptualization as the mind's tendency to seek patterns leads to faulty conclusions. Due to being novices of the discipline, it is likely that students will delve into a subject in an unsystematic, uncomprehending way, resulting in only a superficial grasp of the material, if not just incorrect (Glaser 1991). Ultimately, a "big picture" visual works in conjunction with other materials to ensure that the student has a solid conceptual framework to properly explore the details lurking within the individual facets of this "big picture."

Additionally, the aesthetics of visuals play a role in learning as well. Embellished learning materials aid learning by attracting attention and reinforcing new synaptic connections (Mangurian 2005). Whimsical touches in instructional materials induce emotional arousal which releases chemicals from the limbic system into the brain that support associated synaptic connections. Emotional arousal while processing new material results in students learning more readily and retaining the newly acquired information longer (Leamnson 2000, Mangurian 2005).

As Nilson explains in her book entitled “The Graphic Syllabus and the Outcomes Map”, it is unlikely that students will develop an effective and valid schema for organizing disciplinary knowledge. Unlike those in academia who spend years immersed in a certain discipline or field until such a schema is naturally established in the process, most students will not have enough exposure and length of study of the discipline to develop one. This puts the burden on the instructor to induce a solid understanding of the structure of the discipline onto the students. As Nilson puts it, “it is incumbent on experts - that is, college instructors - to guide the learner through an intellectual apprenticeship.” A visual encompassing the essential concepts and showing the relationships among them facilitates this endeavor. Providing students with an accurate, thought-out structure for making sense of the knowledge may make the difference between whether students retain the knowledge from the course or whether the information will slowly but surely fade away.

Speaking of the issue from the perspective of the teaching of economics, as Moosavian (2016a) explains, “The plurality and variety of concepts, variables, diagrams, and models involved in economics can be a source of confusion for many economics students. However, furnishing students with a visual “big picture” that illustrates the ways through which those numerous, diverse concepts are connected to each other could be an effective solution to clear up the mentioned mental chaos.” He has reviewed the existing literature on the teaching of economics and visualization concluding that “there is enormous potential with visualization to improve the quality of teaching and learning in economics which has not yet been fully employed.” As a practical example, the present paper introduces two wisely designed, visual “big pictures” that can be used as good resources in advanced microeconomics course.[1]

To conclude the literature review, it should be noted that although there is a huge potential with providing visual “big pictures” of complex theoretical economics subjects in clarifying theoretical intricacies, it seems that this potential capacity has not yet been employed fully to clarify theoretical complexities of the economic theory. The present paper is an effort to fill this gap for production theory by visualizing the subtleties and complexities involved in the setting of production theory.

## 3. Main Discussion

Microeconomics consists of two major parts: consumer theory and producer theory. Producer theory involves a wide variety of inherently related concepts, ranging from cost minimization to profit maximization, from market power to market failure, etc. Two of the most fundamental subsections of producer theory are cost minimization and profit maximization. Each of these two processes, *per se*, is an extensive topic in microeconomics, involving many mathematical subtleties and technical details. As such, introducing all the technical aspects of these two processes is beyond the scope of the present paper. However, this section of this paper is to discuss the general relationship between cost-minimization and profit-maximization mostly in a visual way. This section attempts to clarify the theoretical intricacies existing in these two contexts through the use of conceptual visualization. In addition to the elaboration of the general relationship between these two, we will illustrate in what ways some of the major components of each of these two processes are related to those of the other. Finally, we will explicitly introduce some potential applications of the theories of cost and production in the real world.

Many microeconomics instructors prefer to teach producer theory after teaching consumer theory, so that they can draw an analogy between the relationship of utility maximization and expenditure minimization on the one hand and profit maximization and cost minimization on the other hand.[2] Although taking such an approach could be helpful and beneficial in some extent, it is still important to explicitly bring up the subtle distinctions between these two apparently analogous relationships. In other words, although these two pairs of processes are somewhat similar, they have many major differences that need to be cleared up for students to not get confused.

The cost-minimization problem is almost the exact analogue of the expenditure minimization problem, in which the only change is the replacement of the utility function with the production function. The rest of the distinctions are all just notational differences. However, the utility maximization problem differs fundamentally from the profit-maximization problem in many ways. First of all, the utility concept in the utility maximization problem is essentially an ordinal concept, while the production concept in the profit-maximization problem is a cardinal one.

[1] To see a fine macroeconomics example of a visual “big picture”, you can see Zeytoon Nejad (2016d). He provides a visual “big picture” for the course of intermediate macroeconomics.

[2] Moosavian (2016c) introduces and Nejad Moosavian (2016) describes an outstanding example of a comprehensive visual aid elaborating all the components of utility maximization and expenditure minimization processes. It also illustrates visually and mathematically how those components are related to each other.

Another obvious difference is that there is no price choice in consumer theory and price is treated almost always as given while there is a price choice in producer theory in the monopoly and some oligopoly cases. Thirdly, if we think of the inputs quantity choices in producer theory as the counterparts of goods quantity choices in consumer theory, it should be noted that there is no counterpart in consumer theory for the output quantity choice in producer theory. Fourthly, it should be noted that in the wheel of duality visual for consumer theory, the two columns are indeed two alternative approaches to deriving a system of demand functions. However, the two columns introduced in the big picture visual for producer theory are complements to each other, not alternatives, in the sense that for maximizing profit, one first needs to minimize cost and derive the cost function. Finally, the utility maximization problem is a constrained one while the profit-maximization problem is an unconstrained one in conventional microeconomic analysis.

After briefly comparing the wheel of duality in consumer theory to the wheel of relationships in producer theory, we now turn your attention solely to the latter. Now, it will be helpful and informative to examine the relationship between cost-minimization and profit-maximization. Put simply, the cost-minimization process is a general story which is applicable to any number of firms in any market structure, even nonprofits provided that they are interested in operating efficiently under their cost management. However, the profit-maximization process is not a unique one. It is a different story for each type of market structures, i.e. monopoly, imperfect competition, and perfect competition. Figure 1 depicts the spectrum of market structures.

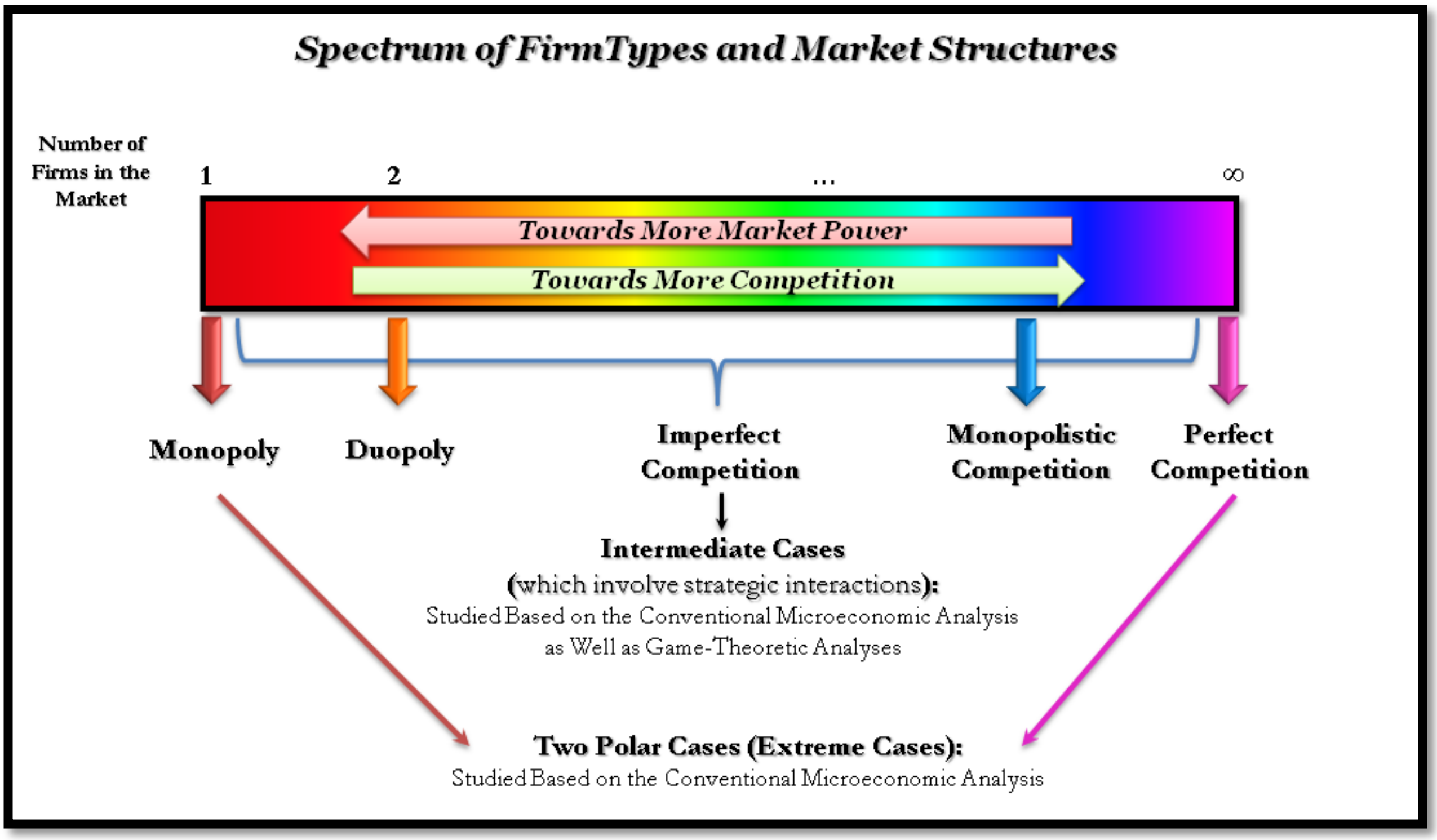

Figure 1. The spectrum of different types of firms and markets

As discussed and shown above, monopoly and perfect competition are two polar cases of market structures, which are usually studied through conventional microeconomic analysis. However, market outcomes become much more tricky and complicated in intermediate cases, i.e. imperfect competition, primarily due to the possibility of strategic interactions amongst firms. Imperfect competition may take many different forms such as oligopoly (few sellers of a product) and monopolistic competition[3] (many sellers producing highly differentiated products), each of which can have different cases, e.g. duopoly is a special case of oligopoly. There are also other forms of imperfect competition such as Monopsony (many sellers but only one buyer) and Oligopsony (many sellers but few buyers), which are hard to be fitted in the above spectrum. After all, in this paper, we will mainly focus on the two extreme cases, leaving the imperfect competition case to be visually decoded in a separate paper in the near future.

[3] It is important to note that although all the market structures illustrated in figure 1 differ primarily in terms of the number of firms in the market, the monopolistic competition has one additional special feature which is a product differentiation within the firms in the market.

The primary goal of the present paper in this section is to illustrate in what ways the components of production theory are related to one another. To do so, we will take advantage of visualization techniques as well as mathematical equations and concepts. The discussion here could be followed by appendix 3, in which two examples are provided for readers to readily grasp the gist of the material being covered.

Now, we start with the cost-minimization part, which is essentially a necessary condition for profit maximization. The main problem to be solved in the cost-minimization process is the minimization of the cost relation (the cost amount) subject to a given production function. By definition, a production function is a mathematical function by which the technology of production within a firm is described. More precisely, a production function is a form of mathematical function that relates the physical quantity of output of a production process to the physical quantities of inputs being used. As Moosavian (2015) elaborates, it is crucial to notice that in specifying a production function, it is assumed that engineering and managerial aspects of technical efficiency has previously been considered, so that the analysis can focus on the problem of allocative efficiency. In fact, this is the reason why the correct definition of production function is given as a relationship between technically "maximum" possible output and the required amount of inputs for producing that output (Shephard, 1970). Despite this, most theoretical and empirical studies define production function carelessly as a technical relationship between output and inputs, and the assumption that such an output needs to be the "maximum" output (and consequently the minimum inputs) quite often remains unspoken (Mishra, 2007). Figure 2 graphically summarizes all the underlying relationships among the cost-minimization problem (including the cost relation, C(W,x), and production function, y=f(x) ), conditional factor demands, $x^c$(W,y), and the cost function, C(W,y).

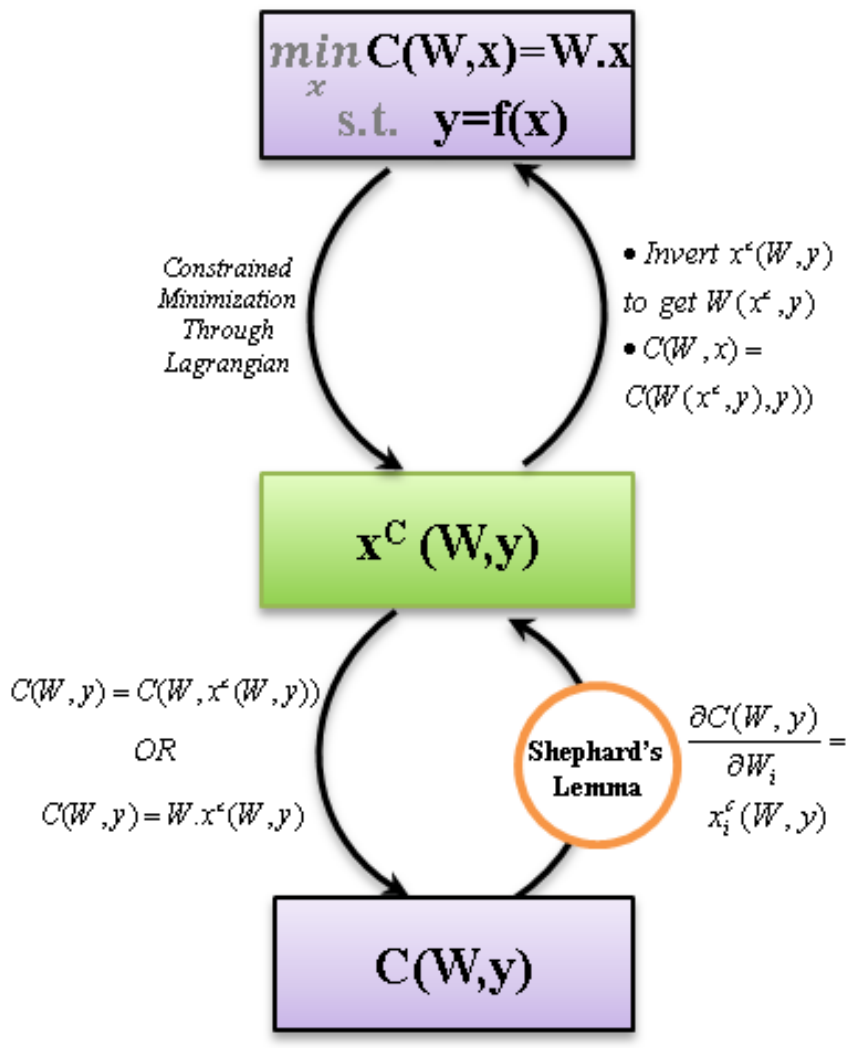

Figure 2. The relationships amongst the cost-minimization problem,

conditional factor demands, and the cost function

Mathematically speaking, the problem located on the top of figure 2 is the generic form of the cost-minimization problem, in which the cost amount relation, as an objective function, is to be minimized subject to a production function of form y = f(x), as a constraint. The first order conditions (FOCs) of this problem will be a vector of conditional factor demands. As explained above, this transition can be made either through the Lagrangian method or through solving a system of equations. Then, if one substitutes the resulting vector of factor demands into the cost amount relation, he or she would easily obtain "the" cost function, which relates the total cost to factor prices given an output level.[4] In order to make the reverse transition to obtain the vector of factor demand from the cost function, one can simply use the Shephard's lemma as outlined above. Additionally, for making a transition between the vector of factor demands to the cost amount relation, one can simply do the mathematical inversion and substitution outlined in figure 2.

The cost function, which is in fact the "total" cost function, is the main outcome of the cost theory. Having characterized the cost function, we can also obtain the average cost function, the average variable cost function, and the marginal cost function through the routine procedures in microeconomics. These three functions will help us identify the Zero-Profit Point and Shutdown Point. These two points provide answers to two of the major questions that every entrepreneur, who is running or planning on operating a firm, especially in a perfectly competitive market, wants to know. Figure 3 shows all the details of the material discussed above in a graphical way.

[4] It is important to clearly distinguish between "the" cost function, and cost amount relation. The cost function is in fact *C(W,y)*, whose arguments are *W* and *y*, while cost amount relation is indeed *C(W,x)*, whose arguments are *W* and *x*.

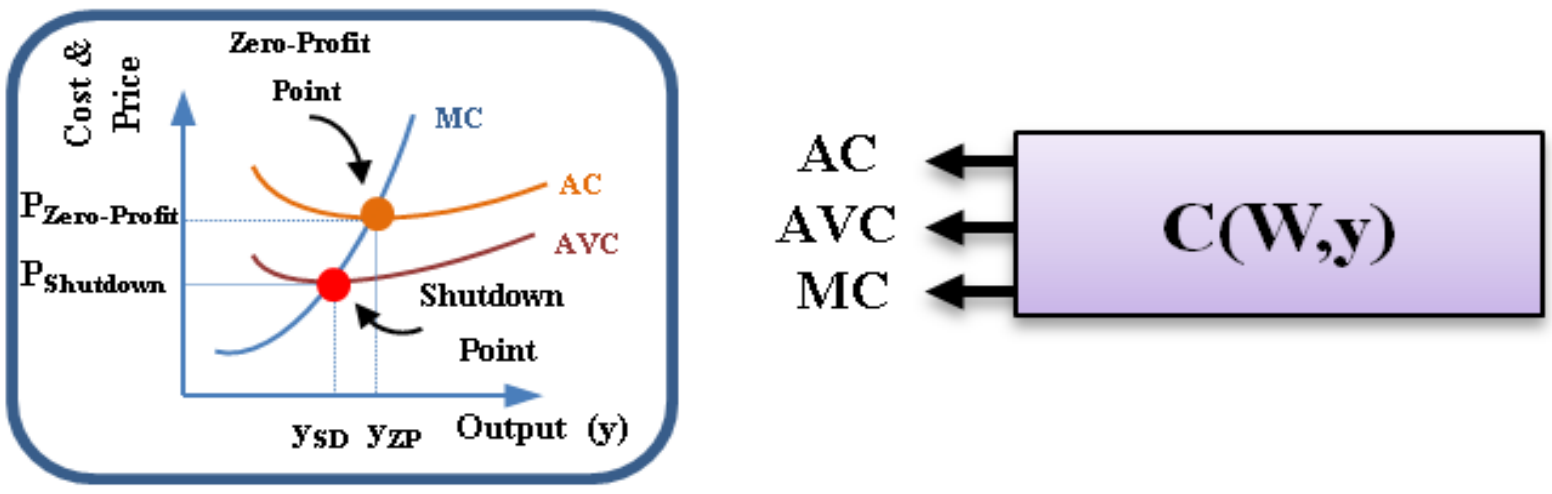

Figure 3. The characterization of Zero-Profit and Shutdown points through the AC, AVC, and MC functions

As mentioned earlier, the cost-minimization problem discussed here is a general one; however, the profit-maximization problem in the perfect competition case slightly differs from that in the monopoly case. Here, we start with explaining the profit-maximization problem of the perfect competition case, and then we will proceed with that of the monopoly case as well.

The main outcome of the cost-minimization problem and the bottom-line of the cost theory, i.e. the cost function, is indeed an input for the profit-maximization problem in production theory. As shown in the following figure, the (total) cost function is subtracted from the total revenue function, and they both together constitute the profit amount relation. This relation has been illustrated through a dashed line connecting the last cell of the left column to the top cell of the right column in figure 7, which depicts a comprehensive visual aid introducing producer theory for the perfect competition case. Figure 4 outlines the profit-maximization problem in a visual way. As shown below, the FOC of the profit-maximization problem is optimal output, aka optimal supply. This optimal output quantity characterizes the optimal scale (i.e. size) of operation. If this optimal output of form y*(P,W) is plugged into the profit amount relation, the resulted equation will be called “the” profit function, which in a sense is the main outcome of the profit-maximization and the bottom-line of production theory.

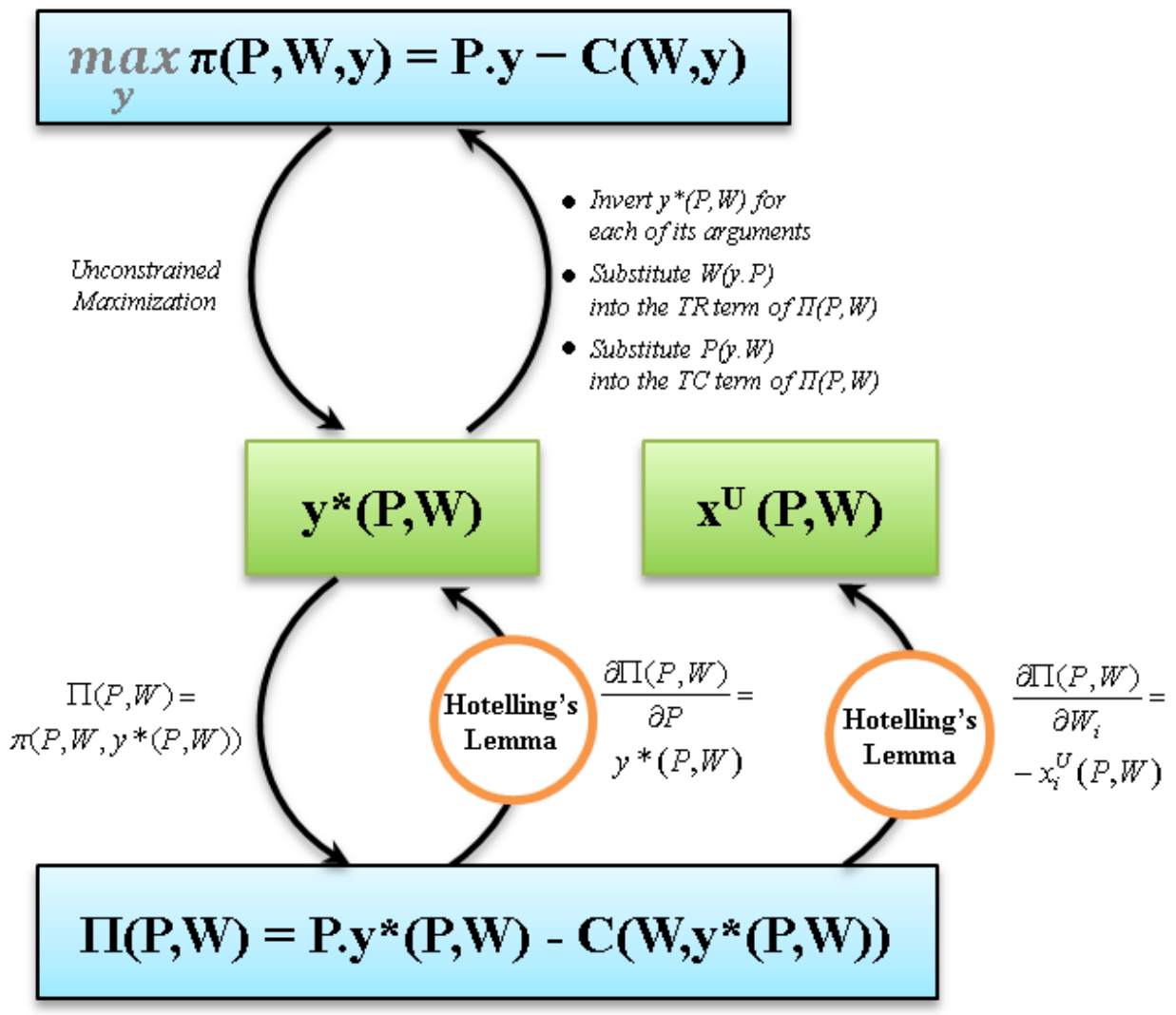

Figure 4. A graphical representation of the profit-maximization problem, production theory, and their major components for the perfect-competition case

In order to make the reverse transition to obtain the optimal output from the profit function, one can simply utilize the left-hand-side Hotelling’s lemma as outlined above. Additionally, for making a transition between the optimal output function to the profit amount relation, one can easily employ the mathematical inversion and substitutions outlined in figure 4. The corresponding graphic representation of the profit-maximization problem for the perfect competition case that results in the procurement of the perfect-competition, optimal output level, y*PC, is illustrated in figure 5.

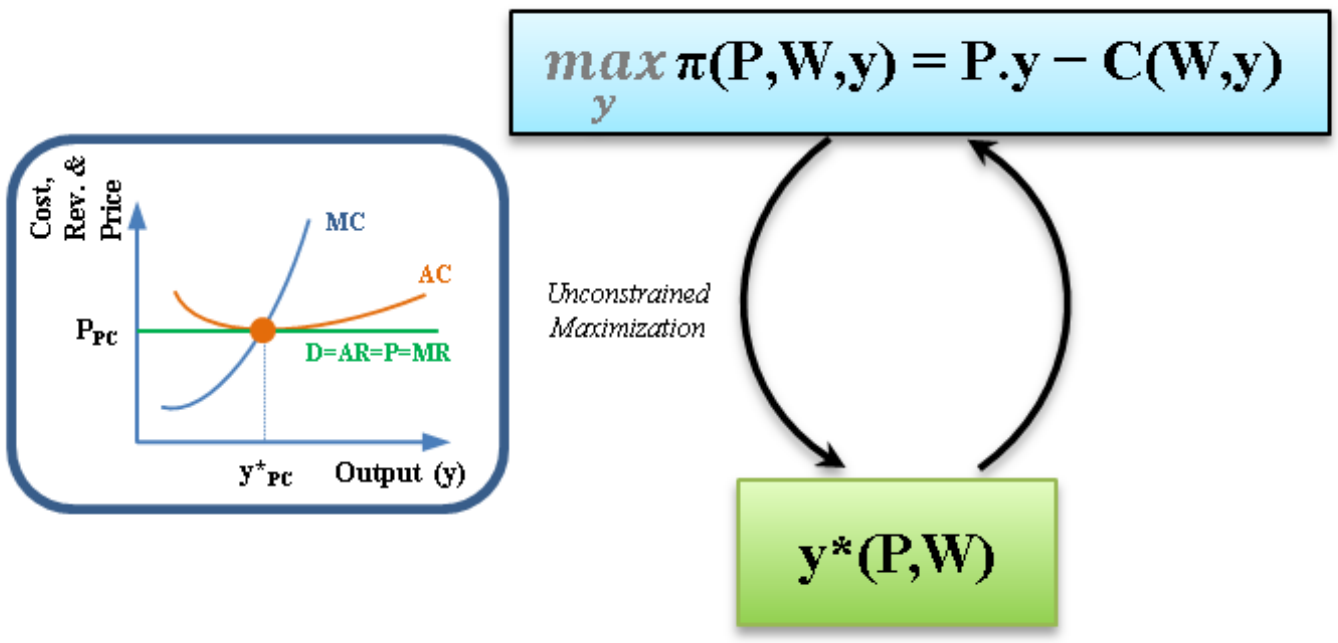

Figure 5. A diagram representing the perfect-competition, optimal output level

One last remaining piece in figure 4 which has not yet been introduced is the vector of unconditional factor demands that is derived by employing Hotelling's lemma, which is introduced in figure 4 (the right-hand-side one). For many, it might be blurred how the conditional factor demands, which are derived in cost theory and either through the cost-minimization problem or through Shephard's lemma, are related to the unconditional factors demands, which are derived in production theory and through Hotelling's lemma. The following picture depicts how these two seemingly unrelated vectors are related to one another. This relationship has been shown in a graphic way as well as a mathematical way in figure 6.

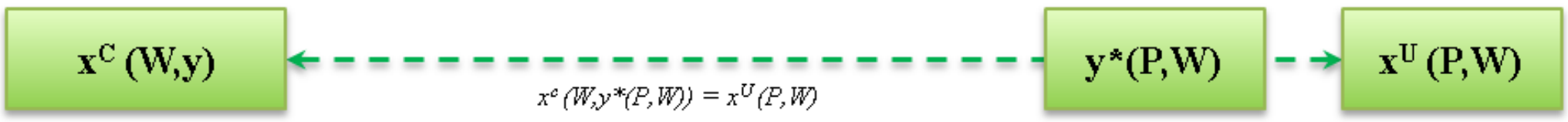

Figure 6. Relationship among conditional factor demand, output supply, and unconditional factor demand functions for the perfect competition case

Now that we have mathematically explained and visually demonstrated how cost theory and production theory as well as their major components are related to one another by using separate figures, it is time to put everything in a single visual aid to show the general picture of how these elements are connected to each other. Figure 7 exhibits the visual "big picture" of cost and production theory for the perfect competition case. The column on left represents the cost-minimization problem, while the column on right represents the profit-maximization problem. This figure includes all the mathematical formulations and operations needed to make theoretically meaningful transitions among different functions existing in the wheel of relationships in cost and production theory.

It is important to note that the cost and production functions are essentially single functions containing all the technological information on the inputs and outputs under study. On the contrary, each of the conditional and unconditional factor demand functions (CFD and UFD) are indeed an extensive form providing possibly a system of equations (i.e. a system of demand functions), each of which represents a demand function for one of the inputs existing in the related production function. The number of equations in each of these systems is equal to the number of inputs in the related production function. Hence, it makes more sense to put a plural "s" at the end of their acronyms as CFDs and UFDs implying that each of these two alone is the representative of a whole system of equations.

Figure 7 summarizes all the operations, equations, and lemmas that help us make the aforementioned transitions. These operations, equations, and lemmas are as follows: Lagrangian method, mathematical substitution, mathematical inversion, Hotelling's lemmas, Shephard's lemma, etc. For a full list of the symbols and notations employed in the visual aids, you can see appendix one. For a larger version of the following figure, you can see appendix 4.

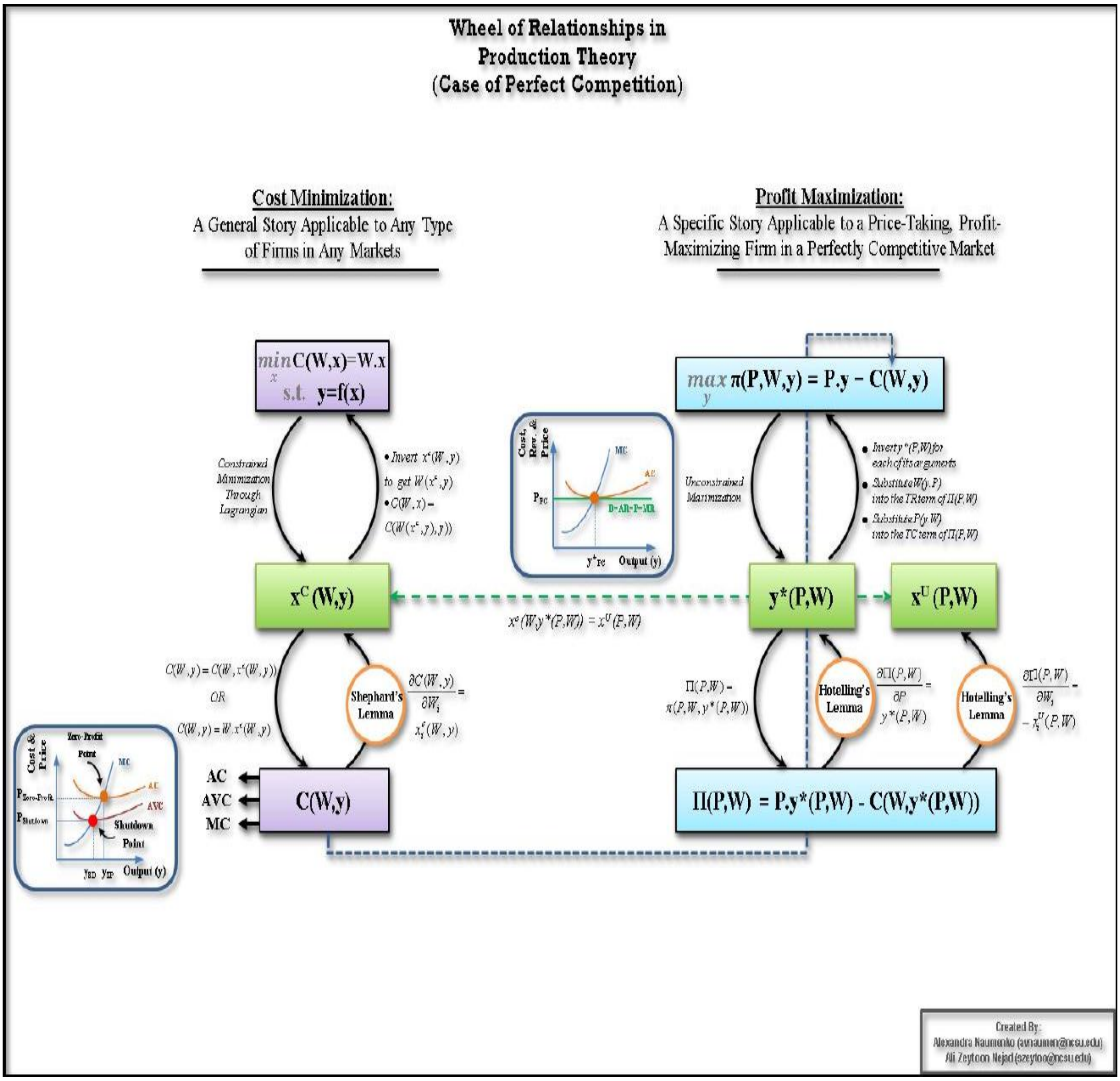

Figure 7. A comprehensive visual wheel of relationships among the cost-minimization problem, the profit-maximization problem, and their main components in the perfect competition case

For the monopoly case, most of the elements and the overall structure remain unchanged, especially for the cost theory part, in which everything is almost the same except that it does not make much sense to define a Shutdown price for a monopolistic firm while such a firm has market power, and as a result, determines prices by itself. Therefore, naturally, such a firm never opts to operate at such a point. Needless to say, such a firm never chooses to operate at the zero-profit point, either; however, since such points are still theoretically possible, we have retained these points in the related diagram.

However, on the right column (the profit-maximization side) many items will change in form for a monopolistic firm compared with those of a perfectly competitive firm, mainly due to the incorporation of a downward-sloping demand curve that a typical monopoly usually faces. The diagram below depicts the right-hand-side column for the monopoly case.

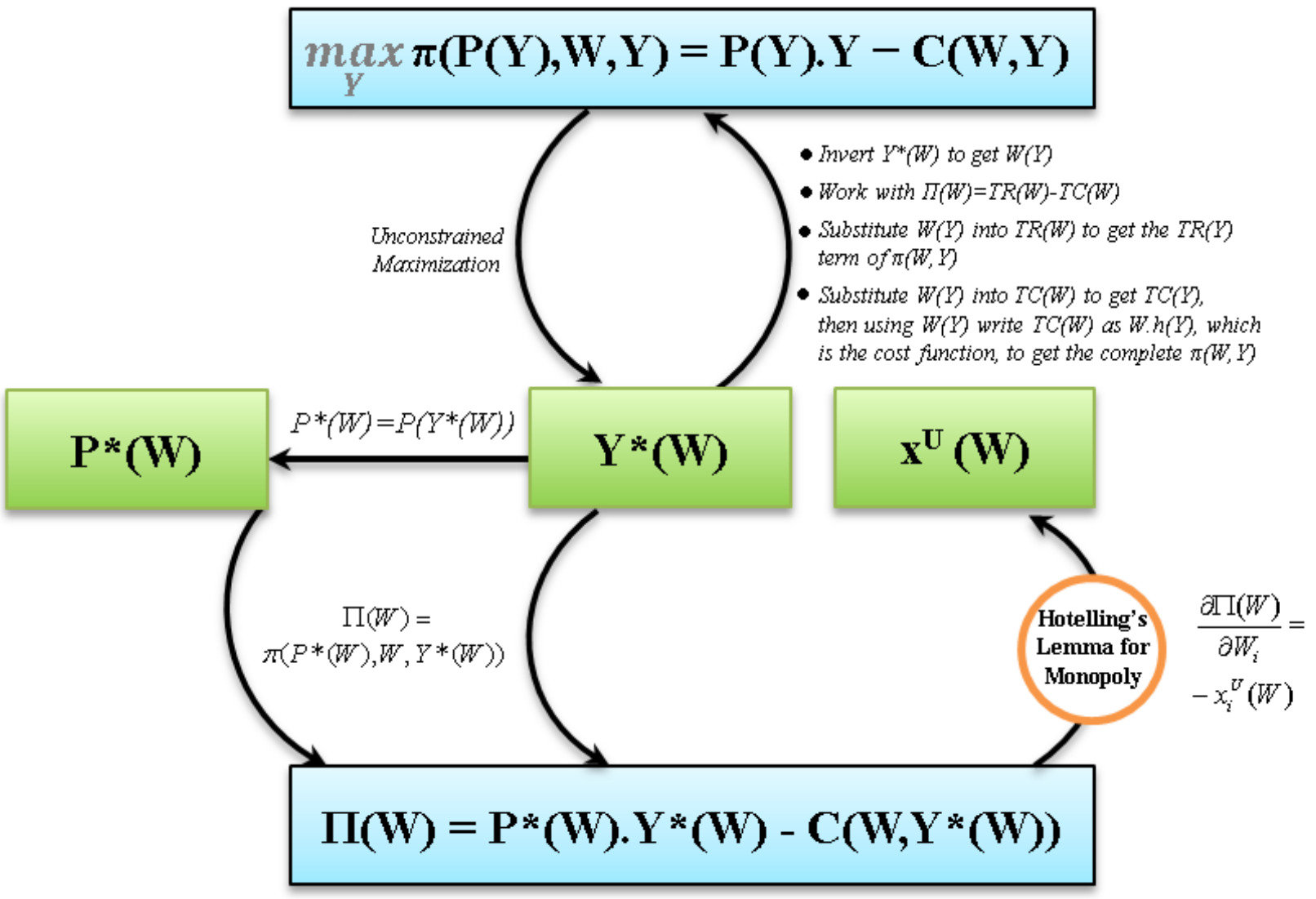

Figure 8. A graphical representation of the profit-maximization problem, production theory and their major components for the monopoly case

As illustrated above, the main difference here is the incorporation of an inverse demand function, P(Y), into the profit amount relation in lieu of the simple market price, P, which we had in the perfect competition case. This added element to the profit amount relation will result in a number of minor changes, which have been demonstrated in figure 8. Other than these minor changes, almost all the rest of the procedures remain similar to those of the perfect-competition case. As an example of these minor changes, the only parameter for deriving the supply, price, unconditional factor demand, and the profit function is solely the vector of input prices, W, as opposed to the perfect competition case, in which we deal with two parameters which are output price, P, and input prices, W, to obtain the mentioned functions. This subtle distinction from the perfect competition results lies in the fact that the output price in the monopoly case is a choice and endogenous variable, not an exogenous one. Another difference here is that as opposed to the perfect-competition case, in which we had a lowercase "y" as a perfectly competitive firm's output, for the monopoly case, we have an uppercase "Y", which represents both the monopolistic firm's output and the market output due to the fact that a monopolistic market is composed of a single monopolistic firm. The corresponding graphic representation of the profit-maximization problem for the monopoly case that results in the procurement of the monopolistic, optimal output level, y*MON, is illustrated in figure 9.

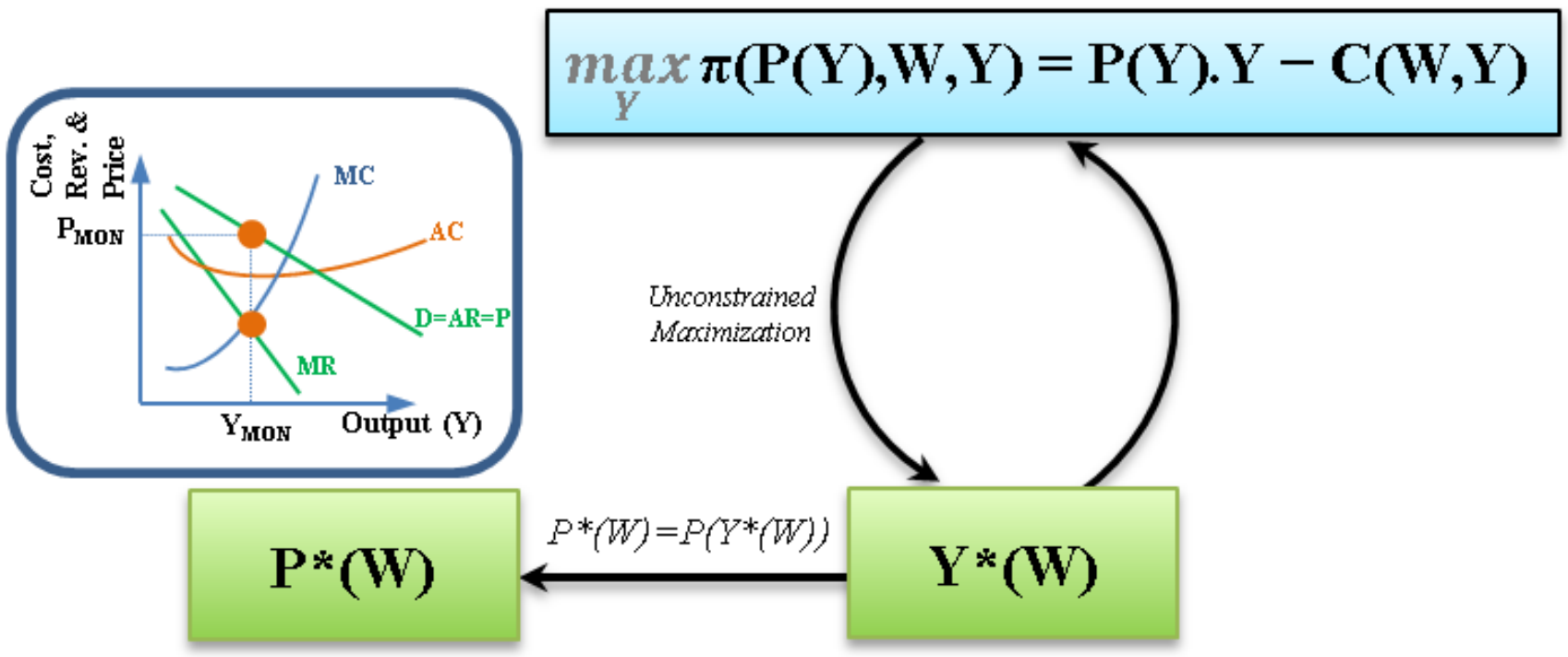

Figure 9. A diagram representing the monopolistic, optimal output level

The following picture depicts how conditional and unconditional factor demand functions (i.e. vectors) are related to one another. This relationship has been shown in a graphic way as well as a mathematical way in figure 10.

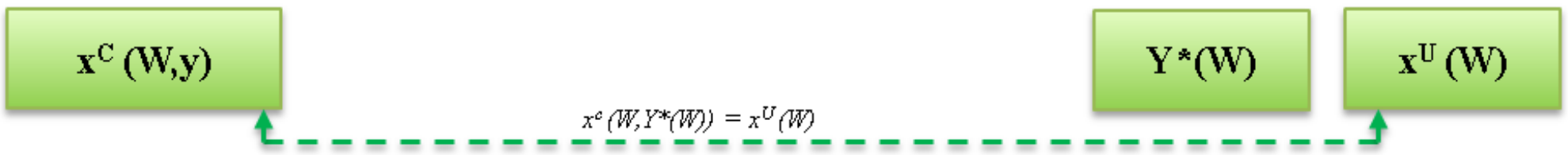

Figure 10. Relationships among the conditional factor demand, output supply, and unconditional factor demand functions in the monopoly case

Figure 11 provides an intuitive illustration of cost and production theory in the modern microeconomics for the monopoly case. It summarizes all the operations, equations, and lemmas that help us make the aforementioned transitions in the monopoly case. Again, for a full list of the symbols and notations employed in the visual aids, you can see appendix one. For a larger version of the following figure, you can see appendix 5.

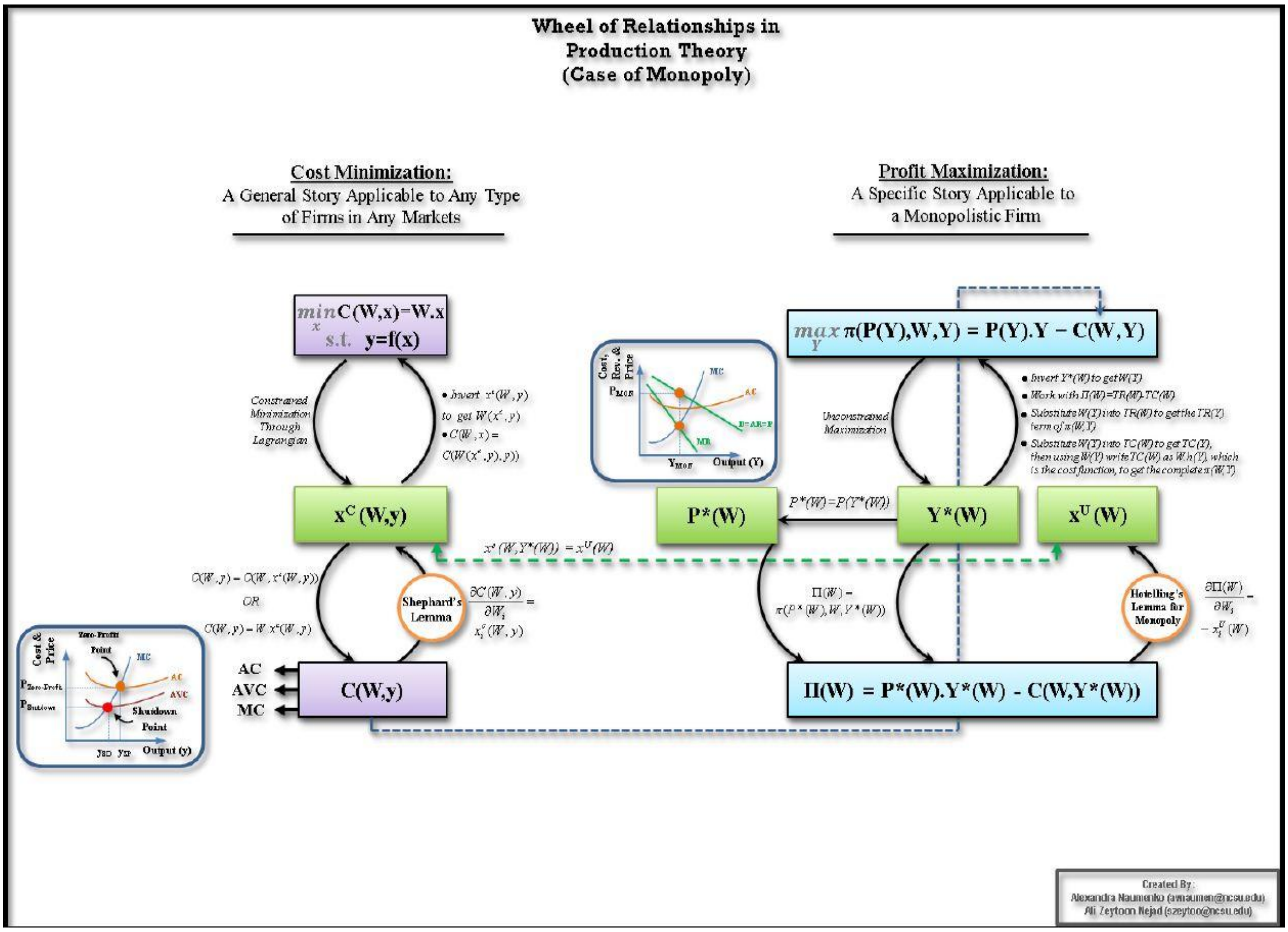

Figure 11. A comprehensive visual wheel of relationships among cost theory, production theory, and their main components in the monopoly case

As you may have noticed thus far, in any system of production, there are principally three major, implicit, and distinct notions of optimality, which are as follows: technical (technological) optimality, input allocative optimality, and output scale optimality. As mentioned earlier, technical optimality is achieved when we specify a correct production function, through which a technological relationship between technically and physically “maximum” possible output and the required amount of inputs for producing that amount of the “maximum” output is identified. Such a mathematical function will give us a “plane” or possibly a hyper-plane, which characterizes a set containing all the possible combinations of inputs employed for obtaining different output levels. Therefore, we achieve technical optimality through the specification of a correctly specified production function. This kind of optimality is formed primarily on the basis of input types (whether they are perfect substitutes, relative substitutes, or perfect complement, etc.) and physical characteristics of inputs and output.

A second notion of optimality which is definable in any production system is input allocative optimality. This sort of optimality is achieved by solving the cost-minimization problem, through which we characterize a set containing the input combinations that are optimal for given output levels. Such optimality will result in a "line" (aka expansion line or path) on the "plane" which we already characterized for technical optimality. This type of optimality is formed primarily on the basis of input prices.

A third notion of optimality that can be defined in any production system is output scale optimality, which in turn implies the scale of inputs employed as well, since output and inputs are linked technologically through a production function. Such optimality will result in a "point" that is located on the "line" which we already characterized for input allocative optimality. This type of optimality is formed primarily on the basis of input and output prices, and in the monopoly case, on the basis of input prices and the output demand function. Figure 12 depicts the three notions of optimality that arise in any system of production. Also, the three optimality concepts have been shown graphically for a typical Cobb-Douglas production function with two inputs (L: Labor and K: Capital).

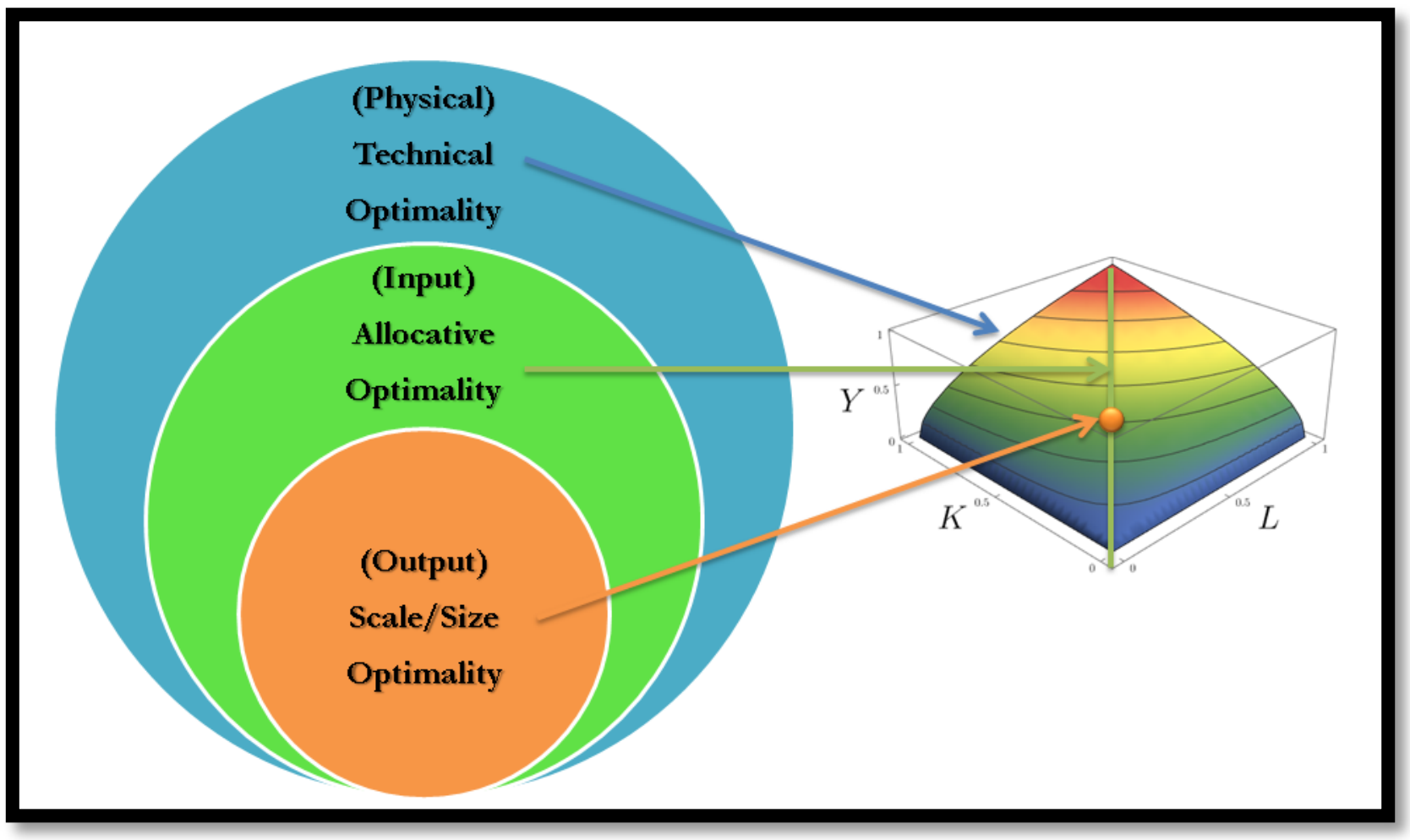

Figure 12. Three notions of optimality existing in any system of production

As you might have noticed thus far, the mathematical and theoretical aspects of the theory of cost and production are so complex in some cases that they hinder us from putting applied aspects of this theory into practice. In addition to this, there have not been sufficiently expository texts elaborating clearly how one can take advantage of those theoretical conclusions in real-life situations. The present paper, which attempts to fill the aforementioned gap, is not only a practical example of how an instructor can clarify theoretical intricacies through the use of conceptual visualization, but it is also an attempt to make clear some real-life applications of cost and production theory. Therefore, we can now take a step forward towards making explicit applications of cost and production theory.

A list of major questions that a typical entrepreneur who is either running or planning on starting a firm may have in mind is as follows:

1- In the short run, at what market price should an entrepreneur stop running the business?

   (A short-run entry/exit question for the perfect competition case)

2- In the long run, at what market price should an entrepreneur stop running the business?

   (A long-run entry/exit question for the perfect competition case)

3- How many units of output should a firm produce and supply?

   (A size or scale question)

4- How many units of inputs will a firm use for the optimal output to be produced?

   (A size or scale question)

5- What price should a monopolistic firm choose? And, when should the price be reset?

   (A price-determination strategy question for the monopoly case)

6- What is the maximum attainable profit for a firm?

   (A profit forecast question applicable to either case)

7- What are the determinants of each of the above-mentioned variables?

   (A question on underlying determining parameters for each of the choice variables)

This is a list of some, but of course not all, of real-world questions in a typical entrepreneur's mind to which cost and production theory provide clear, reasonable answers. If economics students are taught the gist of cost and production theory properly, then they must be able to answer all the questions mentioned above with reference to the economic theory. Suppose an entrepreneur who is evaluating a business opportunity asks an economist these questions. Then, figure 13 will help the economist credibly answer those questions. This figure exhibits the process through which we determine the potential outcomes of the two extreme market structures being examined in this paper, i.e. perfect competition and monopoly.

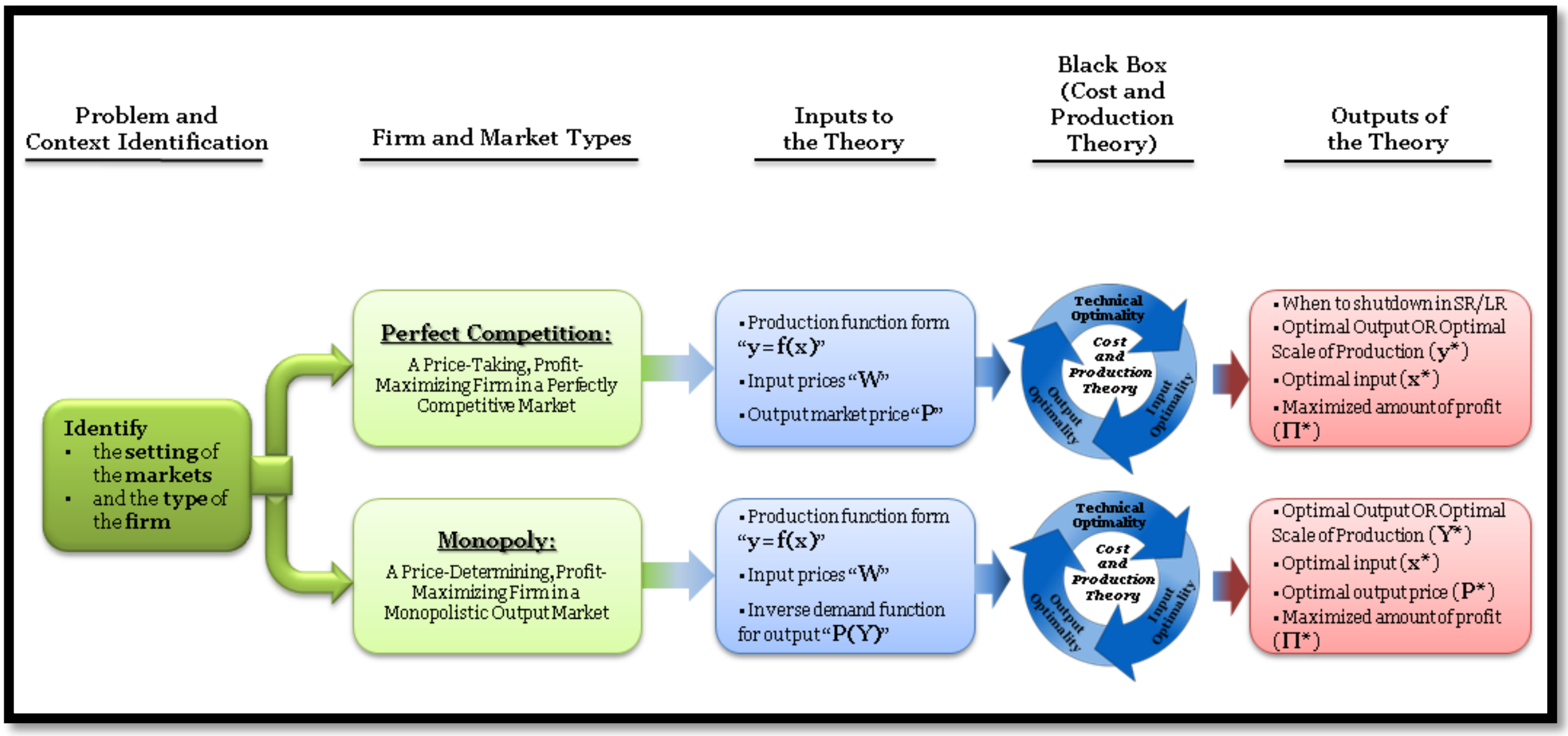

Figure 13. The process of determining the potential market outcomes of the two extreme market structures, i.e. the perfect competition and monopoly

In figure 13, there must be a third branch representing the imperfect-competition market structure, but since such a setting is much different from the two polar cases in terms of both setting and solving procedure, it is beyond the scope of the present paper to discuss that category in detail. In such a setting, we will need to take into account many other factors and elements such as firms' choice variables, the timing of firms' moves, firms' information about market conditions, and firms' information about their competitors' possible actions and payoff profiles. We will discuss this third branch theoretically as well as the best way of teaching it visually, in our opinion, in a separate paper in the near future.

Another noteworthy point to mention here is that for the economic evaluation of many industrial projects, the outputs of cost and production theory must be the inputs of the economic evaluation process of the industrial projects. However, quite often, economic evaluators just simply assume some scale of operation, some sale prices, etc. by using a rule of thumb which, most of the time, are not economically optimal. One possible reason for doing that lies in the fact that they are not familiar with the applied aspects of microeconomic theory, and they feel somehow lost in the theory and as a result cannot figure out how to deal with this complexity. Our work here will help such evaluators more easily figure out how to choose those economic variables and how to proceed with their evaluation in an economically reasonable way. Figure 14 depicts the inputs into and outputs of cost and production theory for real-world applications more compactly than figure 13.

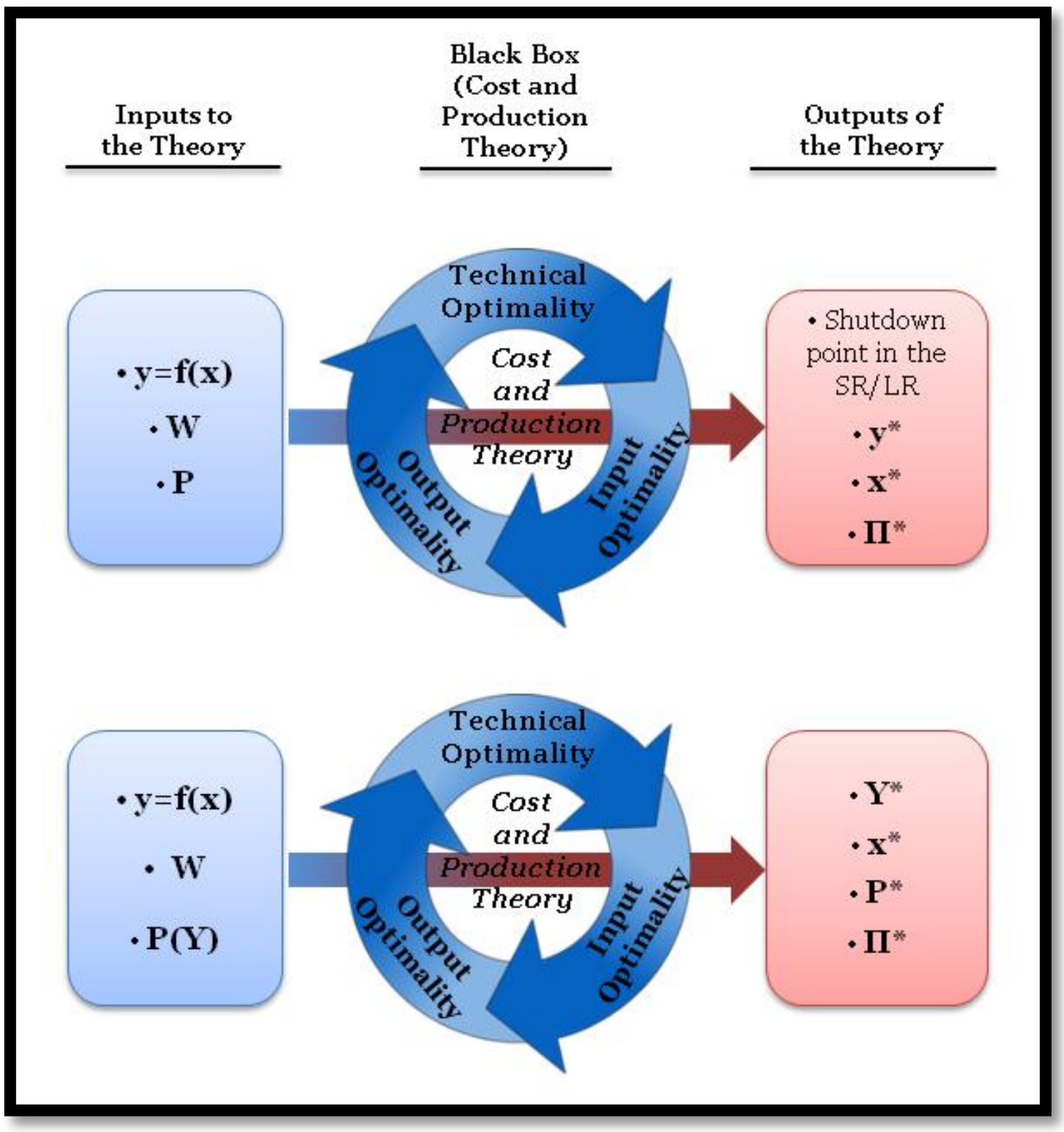

Figure 14. A more compact illustration of the inputs into and outputs of cost and production theory for real-world applications for the two extreme market structures

Figure 14 illustrates the inputs and outputs of cost and production theory in a nutshell to be easily used for real-world applications for the two extreme market structures. The two blue circles in figure 14 are in fact representing the two comprehensive visual wheels of relationships that we already introduced in the paper for the two polar cases of market structures.

In the following, two noteworthy points regarding some of the subtleties of the visual wheel of relationships are raised:

- Alternatively, it is possible to set up the cost-minimization problem using the constraint $D(x, y) = 1$ instead of the constraint $y = f(x)$, in which case we use a distance function to introduce technology to the problem instead of using a production function for that purpose.
- The concepts introduced and discussed in the present paper are all static ones and mostly short-run static ones. We can obviously define and discuss long-run static or even completely dynamic settings and concepts. For instance, we can define dynamic concepts like dynamic price-determination strategy for a monopoly firm, which is all beyond the scope of this paper to be done. However, since the conventional macroeconomic analysis mostly adheres to static settings, we did so, too.

Speaking of pedagogical aspects of the comprehensive embellished wheel of relationships that we introduced in this paper, it must be clear at this point that this graphic provides economics students with a comprehensive visual "big picture" of the relationships among theoretical concepts in production theory. Nilson (2010) points out that "the younger generation of students is not as facile with text as it is with visuals, so a wise idea is to illustrate courses' designs to students so they can 'see' where the course is going in terms of students' learning." Visual aids such as graphic representation of theories, conceptual interrelationships, and knowledge schemata are powerful learning aids because they provide a ready-made, easy-to-process structure for knowledge (Svinicki, 2004; Vekiri, 2002). Nilson (2010) believes instructors should give students the big picture – the overall organization of the course content – very early, and the clearest way to do this is in a graphic syllabus, and that instructors should refer back to the visual big picture to show students how and where specific topics fit into that big picture (Nilson, 2010, p.242).[5] As Moosavian (2016a) puts it, "if we, as instructors, take a course as consisting of three time phases, a big picture can help a class in all the three phases. As he elaborates, "in the first phase, it can be regarded as a graphical outline to illustrate where we are planning to go. In the middle phase, a "big picture" can be treated as a road map or a broad overview of the materials being covered in order to demonstrate exactly what and where in the course the instructor is talking about at the moment. Thirdly, in the final phase, the big picture can be applied as a means of putting things together."

[5] Zeytoon Nejad Moosavian (2016b) takes a step further and elaborates how to provide the big picture of a course to communicate the structure of the course by using an Interactive Graphic syllabus in the context of teaching economics.

## 4. Conclusion

Production theory essentially consists of two simultaneous economic forces: cost minimization and profit maximization, each of which involves deriving numerous functions, such as conditional and unconditional factor demand functions, the cost function, the profit function, and supply function. Nested within the process are many other components such as Shephard's lemma, Hotelling's lemmas, and direct and indirect mathematical relations. The intricacies and hidden underlying influences as well as the nature of having simultaneous economic forces at play all at once make production theory hard for learners to process. This paper aims to offer models that work in conjunction with instructional text to mitigate confusion and improve comprehension. The models introduced in this paper are effective in illustrating the relationship between cost-minimization and profit-maximization, with consideration of technical or physical limitations, which ultimately drive a firm's input and output decisions. These models were constructed according to contemporary learning theories for effective visual techniques and are embedded with mathematical equations and concepts.

The reviewed literature implies that there is a huge potential with providing visual "big pictures" of complex theoretical economics subjects in clarifying theoretical intricacies. It is argued that it is instructors' mission to provide their students with relevant visual big pictures that represent the structure of the associated course with valid, ready-made frameworks for fitting the content. Nonetheless, this potential capacity has not yet been employed fully to clarify theoretical complexities of the economic theory. The present paper is an attempt to close this gap for production theory by visualizing the subtleties and complexities involved in the setting of production theory.

As pointed out earlier, our proposed visual aid presents the comprehensive wheel of relationships for producer theory. As far as we can tell, we are the first to offer such a visual. The graphic offers insight into the two ends of the spectrum of producer theory – perfect competition and monopoly. The paper also evaluated the similarities and differences between perfect competition and monopoly. As far as we are aware, we have not seen the parametric solutions for the monopoly case which we offer in any textbooks. A practical real-world example of how each of the visuals can be used in practice accompanies our theoretical discussion. Embedded throughout the paper are suggestions for how these visuals can help practitioners effectively evaluate industrial or entrepreneurial projects through an economic lens.

In this paper, several comparisons were made, including the wheel of duality in consumer theory versus the wheel of relationships in producer theory, cost-minimization process versus profit-maximization process, and utility-maximization process versus profit-maximization process, and all of their distinctions were explicitly made. In addition, as discussed in the paper, in any system of production, there are principally three major and distinct notions of optimality, which are technical optimality, input allocative optimality, and output scale optimality. These three notions were elaborated to some extent in the paper.

Production theory involves a whole spectrum of market structures, and this paper focused solely on the two extremes – the cases of perfect competition and monopoly. It was beyond the scope of the present paper to discuss the intermediate part of the spectrum in detail. In such a setting, we will need to take into account many other factors and elements such as firms' choice variables, the timing of firms' moves, firms' information about market conditions, and firms' information about their competitors' possible actions and payoff profiles. Future work by the authors will focus on the other market structures in the range between the endpoints of the spectrum, deemed imperfect competition.

## Appendix 1. Symbols and Notations

**max:** Maximize

**min:** Minimize

**s.t.:** Subject to

**x:** Vector of Input Quantities

**W:** Vector of Input Prices

**y:** Firm's Output Quantity (in general and also for a price-taking firm)

**Y:** Market Output Quantity (in general sense and also for a monopolistic firm)

**C(W,x):** Cost Relation (OR, the Amount of Cost)

**y=f(x):** Production Function

**$x^C$ (W,y):** Vector of Conditional Factor Demands (conditional on output)

**C(W,y): "**The" Cost Function (OR, total cost function, i.e. TC)

**AC:** Average Cost Function

**AVC:** Average Variable Cost Function

**MC:** Marginal Cost Function

**Zero-Profit Point:** The point at which AC=MC

**Shutdown Point:** The point at which AVC=MC

**$P_{Zero\text{-}Profit}$:** The price corresponding to the Zero-Profit point

**$P_{Shutdown}$:** The price corresponding to the Shutdown point

**$P_{PC}$:** The optimal price for a perfectly competitive firm

**$P_{MON}$:** The optimal price for a monopolistic firm

**$y_{ZP}$:** The output level corresponding to the Zero-Profit point

**$y_{SD}$:** The output level corresponding to the Shutdown point

**$y^*_{PC}$:** The optimal quantity for a perfectly competitive firm

**$y^*_{MON}$:** The optimal quantity for a monopolistic firm

**π(P,W,y):** Profit Relation for a perfectly competitive firm

**π(P(Y),W,Y):** Profit Relation for a monopolistic firm, which finally reduces to π(W,Y)

**TR:** Total Revenue which equals P.Y for a perfectly competitive firm and P(Y).Y for a monopolistic firm

**AR:** Average Revenue

**MR:** Marginal Revenue

**y*(P,W):** Output Schedule or Supply Function for a perfectly competitive firm

**P:** Output Price for a perfectly competitive firm

**P(Y):** Inverse Demand Function for the output produced by a monopolistic firm

**Y*(W):** Output quantity schedule in terms of input prices for a monopolistic firm

**P*(W):** Output price schedule in terms of input prices for a monopolistic firm

**$X^U$ (W,y):** Unconditional Factor Demand for a perfectly competitive firm

**$X^U$ (W):** Unconditional Factor Demand for a monopolistic firm

**Π(P,W): "**The" Profit Function for a perfectly competitive firm, in which everything is in terms of prices

**Π(W): "**The" Profit Function for a monopolistic firm

## Appendix 2. Mathematical Formulas

Mathematical Formulas in the Cost Theory Part:

Cost minimization problem
$$\begin{cases} \min_{x} \quad C(W,x)=W.x \\ s.t. \quad y=f(x) \end{cases}$$

Substitution needed to obtain the
$$\begin{cases} C(W,y)=C(W,x^c(W,y)) \\ OR \\ C(W,y)=W.x^c(W,y) \end{cases}$$

Shephard's Lemma: $\frac{\partial C(W,y)}{\partial W_i} = x_i^C(W,y)$

Mathematical Formulas in the Production Theory Part – Case of Perfect Competition:

- Profit maximization problem: $\max_{y} \ \pi(P,W,y) = P.y - C(W,y)$
- Substitution needed to obtain the profit function: $\Pi(P,W) = \pi(P,W,y^*(P,W))$
- The profit function: $\Pi(P,W) = P.y^*(P,W) - C(W,y^*(P,W))$
- Hotelling's Lemma 1: $\frac{\partial \Pi(P,W)}{\partial P} = y^*(P,W)$
- Hotelling's Lemma 1: $\frac{\partial \Pi(P,W)}{\partial W_i} = -x_i^U(P,W)$
- Relationship between conditional and unconditional factor demands: $x^c(W,y^*(P,W)) = x^U(P,W)$

Mathematical Formulas in the Production Theory Part – Case of Monopoly:

- Profit maximization problem: $\max_{y} \ \pi(P(Y),W,Y) = P(Y).Y - C(W,Y)$
- Output price schedule in terms of factor prices for a monopolistic firm: $P^*(W) = P(Y^*(W))$
- Substitution needed to obtain the profit function: $\Pi(W) = \pi(P(W),W,Y^*(W))$
- The Profit Function: $\Pi(W) = P(W).Y^*(W) - C(W,Y^*(W))$
- Hotelling's Lemma for the monopoly case: $\frac{\partial \Pi(W)}{\partial W_i} = -x_i^U(W)$
- Relationship between conditional and unconditional factor demands: $x^c(W,Y^*(W)) = x^U(W)$

## Appendix 3. Solved Examples

**Example 1) Problem of a perfectly competitive firm:**

**Setting:**

- A price-taking, profit-maximizing firm in a perfectly competitive market using a single input to produce a single output
- Suppose that the production function is $\mathbf{y=x^{1/2}}$, input price **W**, output price **P**, and output level **y**.

Cost minimization problem
$$\begin{cases} \min_{x} \; C(W,x)=W.x \\ s.t. \quad y=x^{1/2} \end{cases}$$

$x^*(W,y) = y^2$
$C(W,y) = w.y^2$
$\max_{y} \; \pi(P,W,y) = P.y - w.y^2$
$y^*(P,W) = P/2W$
$x^*(P,W) = P^2/2W^2 - P^2/4W^2 = P^2/4W^2$
$\Pi(P,W) = P^2/2W - P^2/4W = P^2/4W$
Therefore, the optimal quantities of output and input will be $\mathbf{y^*(P,W)}$ = P/2W and $\mathbf{x^*(P,W)}$ = $P^2/4W^2$, respectively, and the maximized amount of profit will be $\mathbf{\Pi^*(P,W)}$ = $P^2/4W$.

**Example 2) Problem of a monopolistic firm:**

**Setting:**

- A monopolistic firm using a single input to produce a single output
- Suppose that the production function is **y = x**, inverse demand function for output $\mathbf{P(Y)=Y^{-1/2}}$, input price **W**, and output level **Y**.

Cost minimization problem
$$\begin{cases} \min_{x} \; C(W,x) = W.x \\ s.t. \quad y = x \end{cases}$$

$x^*(W,y) = y$

$C(W,y) = w.y$

$\max_{Y} \; \pi(P(Y),W,Y) = Y^{-1/2}.Y - w.Y$

$Y^*(W) = (2W)^{-2} = 1/4W^2$

$P^*(W) = 2W$

$x^*(W) = 1/4W^2$

$\Pi(W) = 1/2W - 1/4W = 1/4W$

Therefore, the optimal quantities of output and input will be $\mathbf{Y^*(W)}$ = $1/4W^2$ and $\mathbf{x^*(W)}$ = $1/4W^2$, respectively, and the optimal output price to be chosen by the monopolistic firm will be $\mathbf{P^*(W)}$ = 2W, and the maximized amount of profit will be $\mathbf{\Pi^*(W)}$ = 1/4W.

## Appendix 4. A larger version of visual wheel of relationships in the perfect competition case

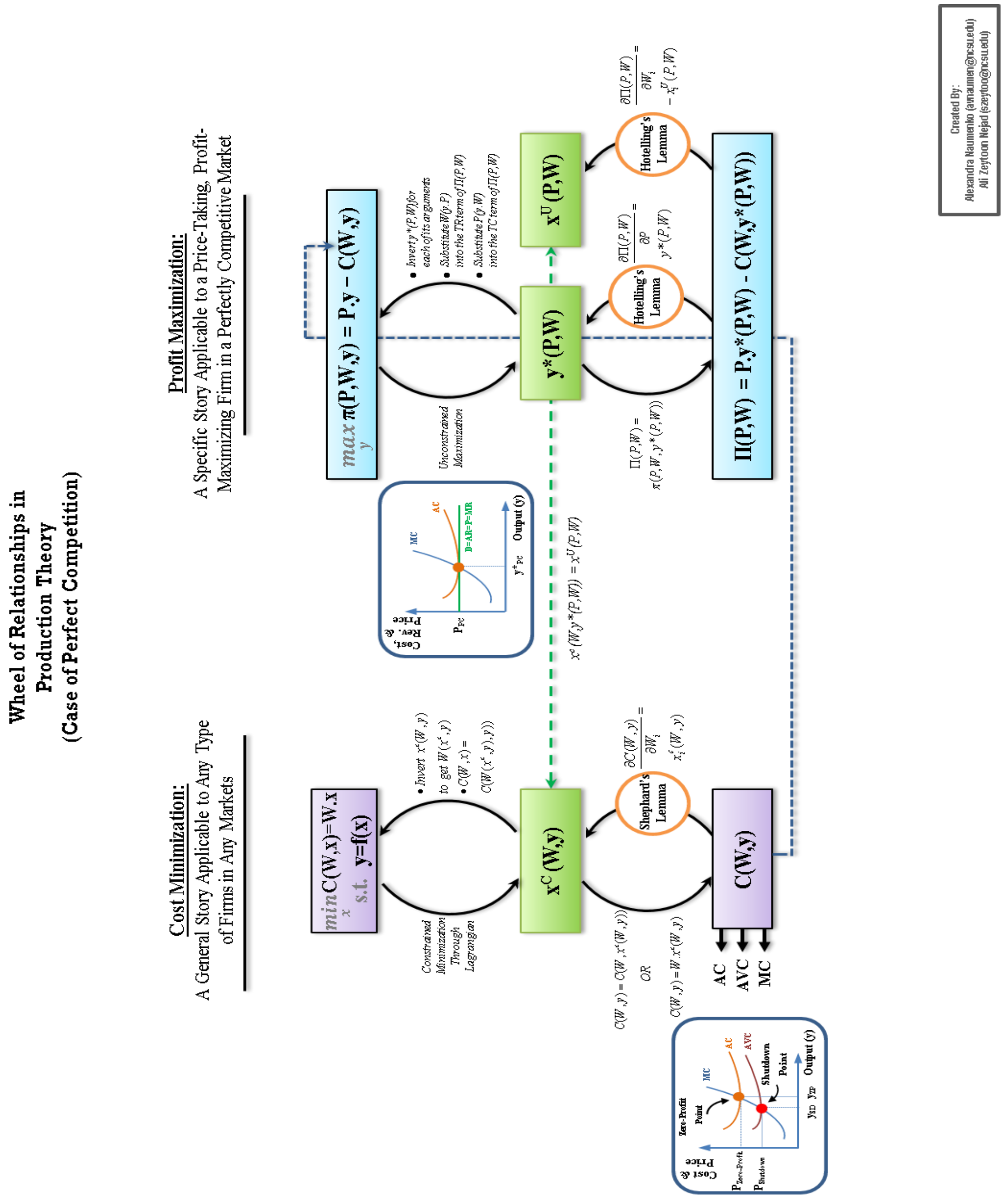

**Appendix 5. A larger version of visual wheel of relationships in the monopoly case**

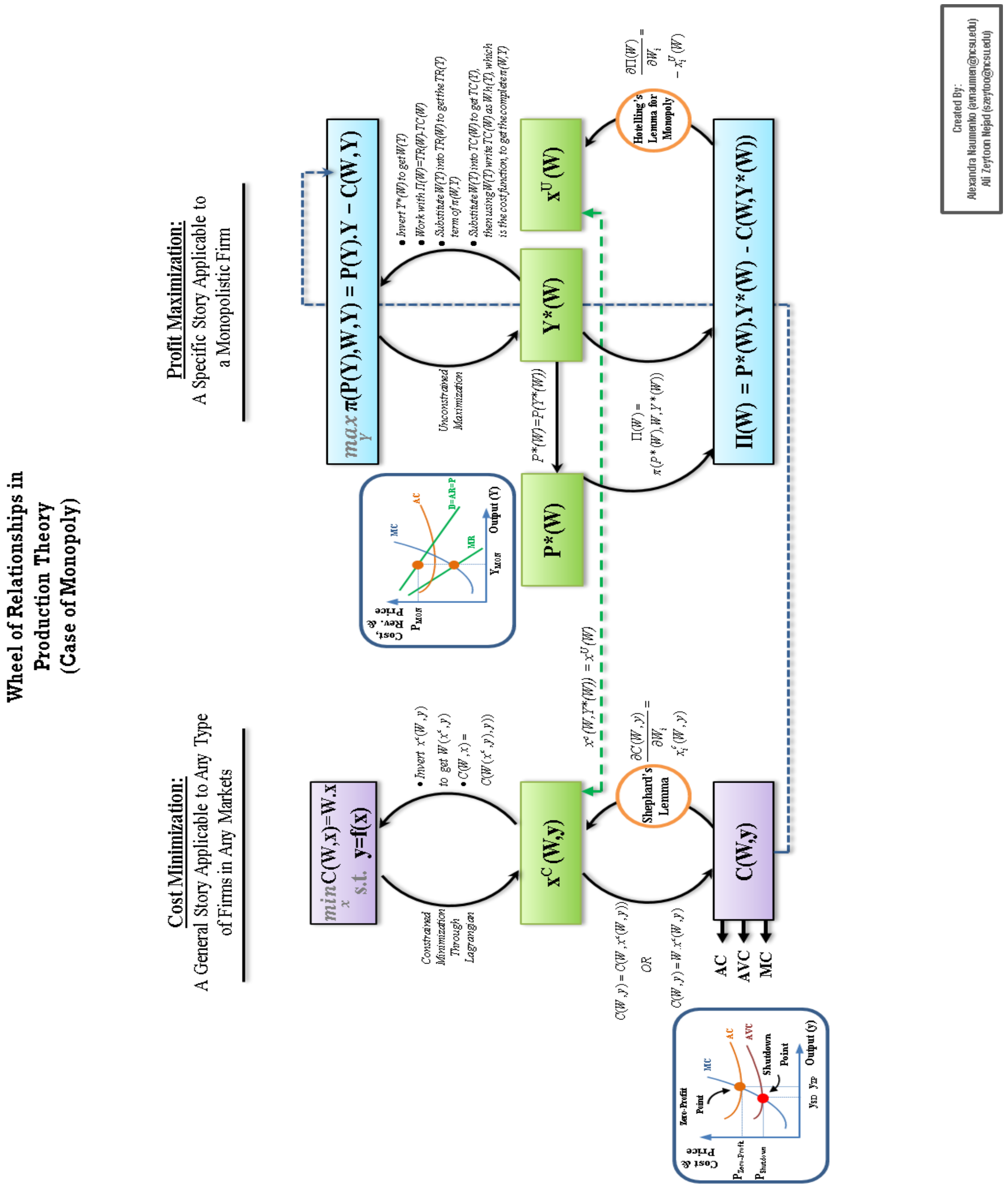